\documentclass[letter,journal]{IEEEtran}

\usepackage[absolute,showboxes]{textpos}

\usepackage{graphicx}
\graphicspath{{figures/}}
\DeclareGraphicsExtensions{.pdf,.jpeg,.png}
\usepackage{cite}
\usepackage{balance}
\usepackage{textcomp}           % More symbols
\usepackage{color}
\usepackage{threeparttable}
\usepackage{float}
\usepackage{booktabs}

\usepackage{varwidth}
\usepackage{fixltx2e}
\usepackage{caption}
\usepackage{subcaption}

\usepackage{pgf}
\usepackage{pgfplots}
\pgfplotsset{compat=newest}
\usepackage{pgfplotstable}
\usetikzlibrary{plotmarks, patterns}
\usepackage{siunitx}

\usepackage{xparse} 
\ExplSyntaxOn
\DeclareExpandableDocumentCommand \mult {  m m  }
 {
  \fp_eval:n { round( #1 * #2  ,2 ) }
 }
\ExplSyntaxOff

\usepackage{tikz}
\usetikzlibrary{arrows.meta,calc,decorations.pathreplacing,fit,shapes.geometric,shapes.multipart,shapes.symbols}

\usepackage{pifont}
\newcommand{\cmark}{\ding{51}}%
\newcommand{\xmark}{\ding{56}}%



\makeatletter
\def\paragraph{\@startsection{paragraph}{4}{1\parindent}{0ex}{0ex}{\normalfont\normalsize\itshape}}%
\makeatother

\title{Security  for the Industrial IoT: The Case for Information-Centric Networking}

\author{
\IEEEauthorblockN{Michael Frey\IEEEauthorrefmark{1},
Cenk G{\"u}ndo\u{g}an\IEEEauthorrefmark{2},
Peter Kietzmann\IEEEauthorrefmark{2},
Martine Lenders\IEEEauthorrefmark{3},
Hauke Petersen\IEEEauthorrefmark{3},
Thomas C. Schmidt\IEEEauthorrefmark{2},
Felix Juraschek\IEEEauthorrefmark{1},
Matthias W\"ahlisch\IEEEauthorrefmark{3}
}

\IEEEauthorblockA{Safety IO\IEEEauthorrefmark{1}
\quad
	HAW Hamburg, Germany\IEEEauthorrefmark{2}
\quad	
	Freie Universit\"at Berlin, Germany\IEEEauthorrefmark{3}
}

\IEEEauthorblockA{\{first.last\}@\{safetyio.com, haw-hamburg.de, fu-berlin.de\},
t.schmidt@haw-hamburg.de}
}

\begin{document}

\IEEEoverridecommandlockouts
%\IEEEpubid{\makebox[\columnwidth]{} \hspace{\columnsep}\makebox[\columnwidth]{ }}
% make the title area

\maketitle

\setlength{\TPHorizModule}{\paperwidth}
\setlength{\TPVertModule}{\paperheight}
\TPMargin{5pt}
\begin{textblock}{0.8}(0.1,0.02)
     \noindent
     \footnotesize
     If you cite this paper, please use the WF-IoT reference:
     M. Frey, C. G{\"u}ndo\u{g}an, P. Kietzmann, M. Lenders, H. Petersen, T.~C.
Schmidt, F. Juraschek, M. W\"ahlisch. Security for the Industrial IoT: The Case for Information-Centric Networking. In \emph{Proc. of IEEE WF-IoT}, IEEE, 2019.
\end{textblock}

\begin{abstract}
Industrial production plants traditionally include sensors for monitoring or documenting processes, and actuators for enabling corrective actions in cases of misconfigurations, failures, or dangerous events. With the advent of the IoT,  embedded controllers link these `things' to local networks that often are of low power wireless kind, and  are interconnected via gateways to some cloud from the global Internet. 
Inter-networked sensors and actuators in the industrial IoT form a critical subsystem while frequently operating under harsh conditions. It is currently under debate how to approach inter-networking of critical industrial components in a safe and secure manner. 

In this paper, we analyze the potentials of ICN for providing a secure and robust networking solution for constrained controllers in industrial safety systems. We showcase hazardous gas sensing in widespread industrial environments, such as refineries, and compare with IP-based approaches such as CoAP and MQTT. Our findings indicate that the content-centric security model, as well as enhanced  DoS resistance are important arguments for deploying Information Centric Networking in a safety-critical industrial IoT. Evaluation of the crypto efforts on the RIOT  operating system for content security  reveal its feasibility for common deployment scenarios. 
\end{abstract}


%\IEEEpeerreviewmaketitle

\vspace{1.25ex}
\section{Introduction}
\label{sec:intro}

Things in the Internet of Things (IoT) are often represented by small embedded controllers which possess orders of magnitude less resources (kBytes of memory, MHz CPU speed, mW of power) than regular Internet nodes, but still need to communicate using  protocols that interoperate in a common infrastructure. One predominant deployment area is industrial automation and surveillance, since embedded controllers are already prevalent in this industry, and adding a networking layer can generate immediate cost and performance benefits for its users. Initial deployments rely on legacy protocols such as MQTT---convergence on a  future common networking standard for the industrial IoT is still under debate.

Today's things are sensors or actuators that speak with a remote cloud or talk with each other locally. The prevalent communication for edge devices happens on wireless channels that are from low power lossy networks (LLNs) in the battery-powered world. Following the IEEE 802.15.4, BLE, or LWPAN standard, these nodes can exchange only small packets at very low rates and sleep frequently. Violating these constraints quickly leads to successive overload, extreme packet losses, and may strongly degrade network operation and node availability. Repeated incidents have shown that the mass of IoT nodes can be both highly threatened and a threat to the global Internet. 

Information Centric Networking (ICN) \cite{adiko-sind-12} was introduced as a  networking paradigm for improved content access in a Future Internet. Ubiquitous caching is a core feature of ICN. NDN (or CCN) \cite{jstp-nnc-09}, its most popular flavor, was designed from a strong security perspective as a pure request-response scheme. 
It became apparent \cite{olg-ccnte-10,bmhsw-icnie-14,sblwy-ndnti-16,szsmb-avdir-17} that ICN exhibits great potential for the IoT. The access of named content instead of distant nodes does not only allow for a much leaner and more robust implementation of a network layer, but in particular the request-response pattern of NDN prevents overloading the receiver with data.

ICN deployment in the IoT has been studied with increasing intensity \cite{bmhsw-icnie-14,pf-britu-15,abcmr-inmcd-16,mwt-tucin-16}, touching various design aspects and practical use cases. Several implementations have become available in common IoT operating systems. CCN-Lite runs on RIOT \cite{bhgws-rotoi-13,bmhsw-icnie-14} and on Contiki \cite{dgv-clfos-04}, NDN has been ported to RIOT \cite{saz-dinps-16}. Thus, grounds are prepared for opening the floor to real-world IoT applications with NDN.

In this paper, we discuss  central security aspects of NDN using the example of an industrial safety system. We introduce a real-world use case which we implemented in a recent prototype and identify key security requirements in Section \ref{sec:usecase}. The fundamental security contributions of the ICN networking layer are derived in Section \ref{sec:security}. Section \ref{sec:eval} is dedicated to comparative analyses of NDN versus traditional IP-based approaches. We further show by measurements that the underlying crypto-complexity can be well handled by constrained IoT nodes. A summary and an outlook conclude this paper in Section \ref{sec:c+o}. 

\section{Use Case: Security and Safety in Hazardous Industrial Environments}
\label{sec:usecase}

\begin{figure*}[t]
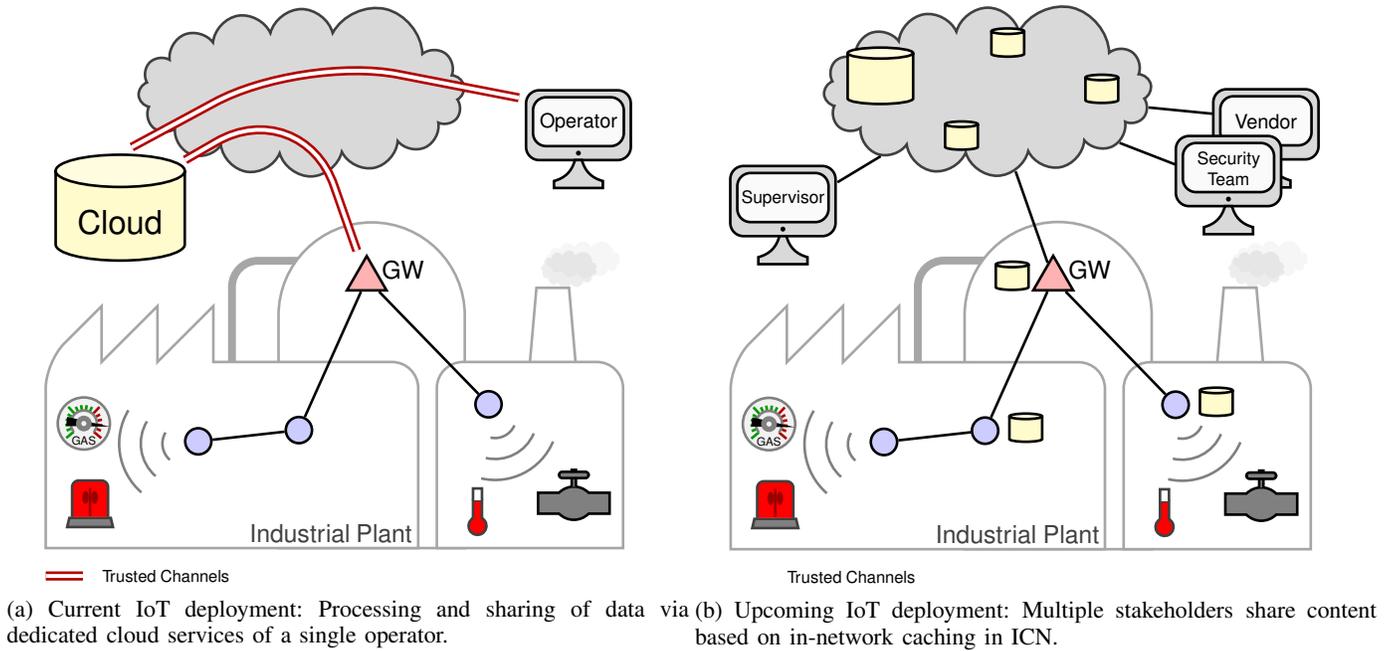

    \begin{subfigure}[c]{0.5\textwidth}
        \centering
        \resizebox{0.90\columnwidth}{!}{\begin{tikzpicture}
    \input{figures/styles.tex}
    \pic[draw=black!35] at (0,0) {scenario=s};
    \pic[scale=0.75,transform shape] at ([xshift=0.6cm,yshift=-0.2cm]s-network.south east) {desktop=operator/Operator/0.8};
    \node[db] (db) at (0.4,1.75) {Cloud};
    \draw[draw=red!70!black,double,shorten <= 2pt,shorten >= 1pt,rounded corners=15pt] (s-gw) -- (s-network.center) -- (db.north east);
    \draw[draw=red!70!black,double,shorten <= 2pt,shorten >= 3pt,rounded corners=15pt] (db.north) -- ([yshift=+0.20cm]s-network.center) -- (operator.north west);
    \draw[draw=red!70!black,double] (0,-0.15cm) -- (0.2cm,-0.15cm);
    \node[anchor=west,font=\sffamily\tiny,scale=0.5,transform shape,inner sep=0pt] at (0.3cm,-0.15cm) {Trusted Channels};
\end{tikzpicture}}
        \caption{Current IoT deployment: Processing and sharing of data via dedicated cloud services of a single operator.}
        \label{fig:deployment_scenario1}
    \end{subfigure}
    \begin{subfigure}[c]{0.5\textwidth}
        \centering
        \resizebox{0.90\columnwidth}{!}{\begin{tikzpicture}
    \input{figures/styles.tex}
    \pic[draw=black!35] at (0,0) {scenario=s};
    \pic[scale=0.75,transform shape] at ([xshift=0.6cm,yshift=-0.2cm]s-network.south east) {desktop=vendor/Vendor/0.8};
    \pic[scale=0.75,transform shape] at ([xshift=0.4cm,yshift=-0.45cm]s-network.south east) {desktop=secteam/Security\\Team/0.7};
    \pic[scale=0.75,transform shape] at ([xshift=-0.75cm,yshift=-0.6cm]s-network.south west) {desktop=supervisor/Supervisor/0.7};
    \draw[draw=black, shorten <= 2pt] (vendor.150) -- (s-network);
    \draw[draw=black,shorten <=2pt] (secteam) -- (s-network);
    \draw[draw=black,shorten <= 2.5pt] (supervisor) -- (s-network);
    \draw[draw=black] (s-gw) -- (s-network);

    \node[db] (db) at (0.8,2.5) {};
    \node[db,scale=0.5,anchor=west,outer sep=3pt] at (s-r4.east) {};
    \node[db,scale=0.5,anchor=west,outer sep=3pt] at (s-r5.east) {};
    \node[db,scale=0.5,anchor=east,outer sep=3pt] at (s-gw.west) {};
    \node[db,scale=0.5,anchor=east,outer sep=3pt] at ([shift=(0:0.75)]s-network) {};
    \node[db,scale=0.5,anchor=east,outer sep=3pt] at ([shift=(45:0.35)]s-network) {};
    \node[db,scale=0.5,anchor=east,outer sep=3pt] at ([shift=(-90:0.25)]s-network) {};
    \draw[draw=none,double] (0,-0.15cm) -- (0.2cm,-0.15cm);
    \node[draw=none,text opacity=0,anchor=west,font=\sffamily\tiny,scale=0.5,transform shape,inner sep=0pt] at (0.3cm,-0.15cm) {Trusted Channels};
\end{tikzpicture}}
        \caption{Upcoming IoT deployment: Multiple stakeholders share content based on in-network caching in ICN.}
        \label{fig:deployment_scenario2}
    \end{subfigure}
    \caption{Current and future deployment scenarios of the industrial Internet.}
\end{figure*}

Industrial safety and control systems are increasingly interconnected to interchange operational conditions locally and to report their status updates to external observers.  A typical deployment scenario consists of  IoT stub networks that are often wireless and confined to the production plant, together with gateways that uplink to an Internet service provider.
Current initial deployment scenarios further involve a (private) cloud  which a dedicated group of trustees can access. Typical stakeholders are the operators of the systems. All parties rely on secure communication channels established between the network endpoints and the cloud. This scenario  builds closed data silos for a preselected, confined group. It is visualized in Figure~\ref{fig:deployment_scenario1}.

Already today it becomes apparent that the number of stakeholders in emerging scenarios will widen---plant operators, emergency teams, equipment vendors, and supervisory authorities may retrieve information about current safety conditions, intermediate operational statistics, as well as long-term reports. Furthermore, even a wider public may legitimately require civil participation in affairs of common impact, as is developing from many open urban sensing initiatives \cite{bjkrb-ssnuc-19}, as well as participatory European laws. Following this demand, data silos need to break up in favour of a flexible, distributed data access that cannot easily rely on preconfigured trusted channels. Still, data might not be uniformly public, but continue to require protection. Protecting the data itself instead of the transmission channels paves the way to transparent data replication and caching---an efficient method for eliding today's silos.
 This heterogeneous environment built from several independent stakeholders is visualized in Figure~\ref{fig:deployment_scenario2}.

Industrial deployments often operate under harsh conditions. In our use case, we consider industrial environments with a threat of hazardous contaminant (e.g., explosive gas) that need continuous monitoring by stationary, as well as mobile sensors. In case of an emergency, immediate actions are required such as issuing local alarms, activating protective shut-downs (e.g., closing valves, halting pumps), initiating a remote recording for first responders and forensic purposes, and eventually may  need to trigger evacuations of the plant or even the region. Such complex settings obviously involve many parties and require a level of robustness which a single uplink to a remote cloud cannot guarantee. 

This use case specifically relies on a fast sensor-actuator network including embedded IoT nodes. The harsh industrial environment raises the challenges of mobile, intermittently connected end nodes, network partitioning, and enhanced reliability from safety requirements.  Devices often need to connect spontaneously, and  a corresponding IoT system  cannot reliably establish  end-to-end channels in many situations. Varying connectivity challenges and mobility, as well as external hazardous impacts are much easier mitigated in a replicative environment, where data diffuses hop-wise in an asynchronous fashion. It is easy to build such a compliant networking layer based on NDN primitives \cite{gksw-hrrpi-18}.

Typical industrial plants are widespread with sparse network coverage, so that mobile workers or machines face intermittent connectivity at scattered gateways. Some sensors and actuators are infrastructure bound, others are independent, battery-powered embedded devices (e.g., body equipment). Such devices are susceptible to battery drains and can process only a few packets per minute on average. They are easily challenged by various distributed denial-of-service (DDoS) attacks. Hence networking approaches should minimize the DDoS surface and protect the embedded edge components.

Taken from real-world deployment, this study makes the case for a distributed, multi-stakeholder environment and identifies three major objectives for the networking layer:

\begin{enumerate}
	\item Allow for ubiquitous multiparty data access without pre-established secure data channels or VPNs in the constrained IoT.  
	\item Provide a robustly secure networking infrastructure that is resilient to varying link conditions and mobility with the ability to recover locally from intermittent impairments.
	\item Raise the barriers for DDoS attacks of constrained devices and confine the attack surface of unwanted traffic to local links.
\end{enumerate}

We will show in the following, how the NDN approaches to Information Centric Networking can significantly contribute to these goals. We will also assess the shortcomings of current IoT solutions such as MQTT~\cite{mqtt311} and CoAP~\cite{RFC-7252}.

\section{Security Contributions of NDN}
\label{sec:security}

According to our use case, an industrial IoT deployment enhances requirements in the security and safety domain, but on the other hand narrows the utilization of ICN functions down to rather specific settings. In this section, we will discuss the three security aspects derived from our use case and identify certain benefits for NDN from its specific deployment in an industrial setting.

\subsection{Ubiquitous data access in the constrained IoT}

Sensor data need to be accessible both in the local constrained IoT, and in the remote for various stakeholders. Safety and security of the industrial monitoring system indeed largely depend on its availability even under the harsh conditions of local or regional incidents with intermittent connectivity. As critical industrial facilities are always also susceptible to malicious threats, utmost resilience against (networked) attacks is strongly desirable. Clearly, a centralized cloud-based approach falls short as tampering the cloud has proven to be a pronounced attack vector (cf. the Cloudflare attack 2013).

Ubiquitous caching is the most striking contribution ICN makes to the security and safety of the distributed information system. Configuring the constrained nodes as well as the gateway to replicate and store IoT data for (most of) its lifetime will maximize redundancy and minimize unavailability of critical information. It is noteworthy that common IoT data is small and of limited lifetime---archives being a well-localized exception. Furthermore, flash storage in constrained nodes is the least scarce resource and typically can accommodate an `infinite' amount of IoT data. 

Local mass storage facilitates the DTN nature of ICN for the IoT. The hop-by-hop transmission of sensor readings and actuator commands  increases resilience in the presence of caching. When links re-establish after mobility handovers or failures, the NDN network layer can easily resume the content propagation and will thus provide an efficient self-healing mechanism. 

\subsection{Robustly secure networking infrastructure}

Sensors and actuators of the constrained IoT are typically challenged by maintaining an authenticated or even encrypted data channel to some remote data repository. In addition, unstable and lossy links in IoT edge networks make it hard to persist a stateful communication relation. Also for these reasons, IoT nodes are commonly deployed behind gateways that execute protocol translations (e.g., DTLS versus TLS) and thereby intercept secured channels. This sacrifices end-to-end transport security and exposes a significant attack surface at the gateway.

By authenticating or encrypting content instead of channels NDN circumvents these operational challenges of the IoT. As each content chunk can be hopwise replicated throughout the network without impairing its security measures, data integrity and confidentiality remain independent of transport or paths. Moreover, there is no requirement of performing synchronous actions between specific endpoints on the Internet which makes the security layer robust against link failures and network disconnects.   

\subsection{DDoS resistance}

Constrained nodes on the low power lossy wireless are easy victims of resource exhaustion when receiving too many IP packets. A gateway may commonly shield the IoT nodes from the global Internet and may even perform some (general) rate limiting, but it cannot reasonably track individual resources of nodes nor hinder the communication needs of the application use case.
In addition, a malicious member of the IoT stub domain may not only jam radio channels, but utilize IP multihop forwarding to overload remote nodes. Conversely, as has been recently reported from the MIRAI incident, huge multiplicities make IoT nodes an interesting amplification tool for attackers.  

A key design objective of ICN had been the reduction of this IP attack surface with respect to distributed denial of service attacks. In NDN this led to designing a request-response communication scheme without node addresses that hinders the plain transmission of unwanted content to a receiver. For a few years, it was the believe that NDN can be DDoS resistant by design, until Interest-  and state-based attacks were discovered \cite{wsv-bipmc-12}. Subsequent work \cite{gtuz-ddndn-13,wsv-bdpts-13} elaborated the threats of Interest flooding and overloading FIB and PIT structures by user-generated names and content requests. This has proven difficult to mitigate in general \cite{sws-rcani-15}. However, in a specific industrial setting of pure machine-to-machine communication with well known traffic patterns, buffers and PIT tables can be  pre-configured according to well-formed communication flows. Hence, Interest flooding can be detected at the first hop and eliminated by the receiving stack (e.g., by hitting PIT limits). State-based attacks can thus be restricted to the local link which can never be protected by a network layer.

\section{Comparative Assessment}
\label{sec:eval}

We are now ready to a qualitative security comparison of our ICN solution with the common IP-based protocols MQTT and CoAP. We also evaluate the complexity of content object security that is inherent to ICN, but for a quantitative performance analysis we refer to \cite{gklp-ncmcm-18}.

\subsection{MQTT}

MQTT is a message-based publish subscribe protocol, with a special focus on low bandwidth environments.
A typical MQTT network involves a client that publishes data on a specific \emph{topic}.
Each topic is managed by a server (or \emph{broker}) which distributes data about the topic to subscribers.
By default, a message that has been published and distributed to the consumers by the broker is deleted after delivery.
Different QoS levels allow for storing messages on the broker or advanced reliability on top of the transport protocol.

Low-end IoT devices are challenged by basic MQTT, as MQTT communicates over TCP.
A lightweight version of MQTT is provided by \emph{MQTT for Sensor Networks} (MQTT-SN) \cite{mqttsn12}.
MQTT-SN is tailored to wireless domains and optimized for devices that are constrained in energy, processing, or storage.
It is implemented on top of UDP and replaces topic strings by topic IDs to shorten messages.

In MQTT as well as MQTT-SN, security features depend on the broker implementation.
Using username and password, or alternatively a client certificate, the broker may authenticate the client it connects to.
If TLS (or DTLS) is used, the client may also authenticate the server.
However, there is no end-to-end security support between publisher and subscriber.
This threatens message integrity when the broker changes content, because subscribers do not have an out of the box mechanism to verify the content.
To protect the payload, additional encryption efforts of application data are required on top of MQTT.

In general, MQTT assumes a trust relationship between broker, publishers, and subscribers.
Usually, authentication and authorization is ignored completely, to simplify device management.
This trust assumption reflects current deployment models, in which either brokers and clients are under the same administrative control, or where service contracts between end devices and a cloud network with broker service exist.

\subsection{CoAP}

The Constrained Application Protocol (CoAP) is standardized in the IETF with the aim for replacing HTTP in constrained deployment scenarios.
CoAP operates on top of UDP and defines a compact protocol header.
It specifies three communication schemes: (i) polling, (ii) push, and (iii) observe.
Using push and observe, CoAP implements publish subscribe scenarios.
In contrast to push, observe does not require explicit subscription in advance but delivers data to clients based on pre-configuration at the server side.

To enable M2M communication, CoAP implementations usually provide both client and server capabilities.
Thus, without an explicit intermediary node such as a broker in MQTT, CoAP nodes may interact directly with each other.

The security support in CoAP is more advanced compared to MQTT, even though several specifications are still under discussion in the IETF.
CoAP is secured on the transport layer using DTLS or alternatively on the application layer using specific extensions such as OSCoAP, which allows for object security in CoAP.
However, it is worth noting that DTLS might conflict with constrained environments as packet sizes increase.
On the other hand, current approaches for object security may conflict with privacy as not all CoAP headers are encrypted and, for example, may reveal content names.

\begin{table}
\center
\label{tbl:comparison}
\caption{Comparison of MQTT, CoAP, and ICN with respect to security measures.}
\begin{tabular}{lccc}
\toprule
 & MQTT & CoAP & ICN \\
 \midrule
Ubiquitous caching & \xmark & (\cmark) & \cmark
\\
Object security & \xmark & (\cmark) & \cmark
\\
Name privacy & \xmark & (\cmark) & (\cmark)
\\
Infrastructure protection & (\cmark) & \xmark & (\cmark)
\\
End node protection & \xmark & \xmark & \cmark
\\
\bottomrule
\end{tabular}
\end{table}

\subsection{Comparing MQTT, CoAP, and ICN}

\paragraph*{Caching}
Caching does not only improve performance in terms of faster data delivery but also increases data availability and robustness.
A common malicious scenario includes a denial of service attack.
With proper replication, the origin data source can go offline without loosing data in the global network.
MQTT is easily threatened by this kind of attack because of the dedicated broker service.
CoAP inherently supports caching on intermediary nodes.
However, this mitigation is only implemented on the application layer.
In common single stakeholder scenarios, where CoAP servers are managed by a single administrative domain, this usually does not help, in particular when network providers are under attack.
ICN provides ubiquitous in-network caching that is independent of individual stakeholders.
Thus, attacking a specific content source is intricate.

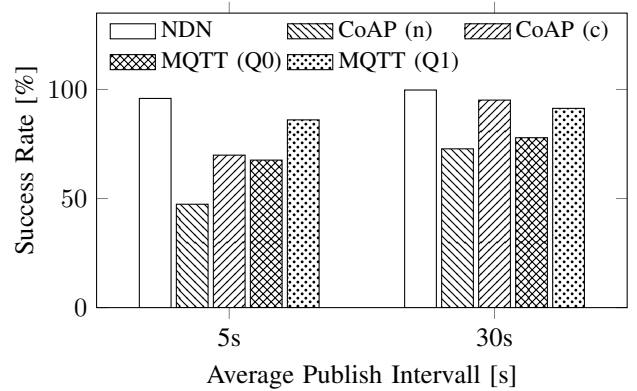
\begin{figure}[b]
  \centering
  \begin{tikzpicture}[baseline]
  \begin{axis}[
    ybar, area legend,
    width=0.975\columnwidth, height=5.5cm,
    bar width=12pt,
    ymin=0, ymax=135,
    ylabel={Success Rate [\%]},
    ylabel shift = -3pt,
    xlabel={Average Publish Intervall [s]},
    xmin=0.5,xmax=2.5,
    xtick=data,
    xticklabels={5s,30s},
    legend cell align=left,
    legend style={draw=none,legend columns=3,anchor=north,at={(0.5,0.990)},/tikz/every even column/.append style={column sep=0pt},/tikz/every odd column/.append style={column sep=0pt},inner sep=0pt,outer sep=0pt,font=\small},
    ]
    \addplot[fill=white] coordinates {
      (1,95.88749200937566) (2,99.74468085106383)
    };

    \addplot[pattern=north west lines] coordinates {
      (1,47.391390597103744) (2,72.78129952456418)
    };
    \addplot[pattern=north east lines] coordinates {
      (1,69.89443767762882) (2,95.12340764331202)
    };

    \addplot[pattern=crosshatch] coordinates {
      (1,67.62) (2,77.88235294117647)
    };
    \addplot[pattern=crosshatch dots] coordinates {
      (1,86.06) (2,91.34)
    };

    \legend{NDN, CoAP (n), CoAP (c), MQTT (Q0), MQTT (Q1)}
  \end{axis}
\end{tikzpicture}%
  \caption{Resilience of NDN vs. CoAP vs. MQTT.}%
  \label{fig:reliability}
\end{figure}

\paragraph*{Reliability} IoT nodes connected via low-power wireless networks suffer severely from lossy communication channels. Even the transmission of small data chunks  to the gateway is frequently impaired by unstable links, and transport protocols are challenged to  cope with the unstable environment in a reliable fashion. We compare NDN, confirmable and non-confirmable CoAP (c/n), and MQTT (Q0/Q1) in Figure~\ref{fig:reliability}. The success rate of packet delivery was measured in two large experiments of 50 nodes from the FIT IoT testbed at different publishing intervals. Low power lossy radios of the IEEE 802.15.4 standard were deployed with link-layer retransmissions set to four. Results clearly demonstrate the superior reliability of the hop-by-hop approach of NDN, while even the reliable variants of CoAP (c) and MQTT (Q1) fail significantly by 30 \% resp. 15 \% in the tighter scenario of publishing every $5~s$. NDN always delivers more than 95~\% of the packets, the success rate approaching 99.9~\% in the more relaxed publishing at  $30~s$.

\paragraph*{Object security}

Security of content objects is crucial in inter-domain scenarios, in particular in the industrial Internet where sensors communicate sensitive information or actuators interact with critical infrastructure components based on data.
Ideally, content can be forwarded by any node in the network without sacrificing security.
MQTT and CoAP need additional efforts to achieve this objective.
ICN, on the other hand, has been designed with democratized content distribution in mind.
In-network caching is not limited to specific service nodes but envisioned to run on any network node that is willing to share resources for caching.
Consequently, content security is a first principle in ICN, allowing multi-stakeholder scenarios with respect to scalable and secure content distribution.
In ICN, trust is not based on contracts but technically provided by design.

\paragraph*{Infrastructure protection}

 CoAP runs on top of UDP.
As UDP is a connection-less protocol without congestion control, it can easily operate IP packet bursts and spoofing.
Having IP spoofing in place, an attacker can initiate a reflective amplification attack, in which the attacker sends a small request towards the CoAP server that replies with a significantly larger packet to the victim (i.e., the spoofed IP address).
Amplification attacks are common in the current Internet and a major threat for operators.
With increased deployment of CoAP, we will experience more of such attacks in the future.

MQTT makes spoofing attacks much more challenging because of TCP.
However, in MQTT-SN, TCP is replaced by UDP to reduce overhead on low-end IoT devices and thus opens up the identical attack surface. 
On the contrary, ICN abandons the end-to-end paradigm completely and provides de-localized services off the shelf.

\paragraph*{End node protection}

 End nodes are not protected in MQTT and CoAP but may receive arbitrary amounts of unwanted data.
Security extensions may enable authentication and authorization but protection against unsolicited traffic requires firewall extensions, either as infrastructure middleboxes, or  as dedicated local software component running on the end node.
The latter conflicts with constrained resources of low-end IoT devices.
An industrial Internet benefits from ICN as ICN does not support end-to-end communication. 
It thus protects end devices against malicious traffic without additional overhead.

\paragraph*{Name privacy}

To comply with privacy requirements, obfuscating the requested content name in the content delivery infrastructure is important. 
Implementing this with low overhead and strong privacy protection is one of the most challenging tasks in content delivery scenarios, yet.
Neither MQTT, nor CoAP, nor ICN provide a solution out of the box until now.
The hope here is that the ICN community will introduce a sufficient solution in the long-term because naming is a key component, which affects all applications on top of an ICN network layer.

\subsection{Expenses of content security in ICN}

\begin{figure}
    \resizebox{\columnwidth}{!}{\input{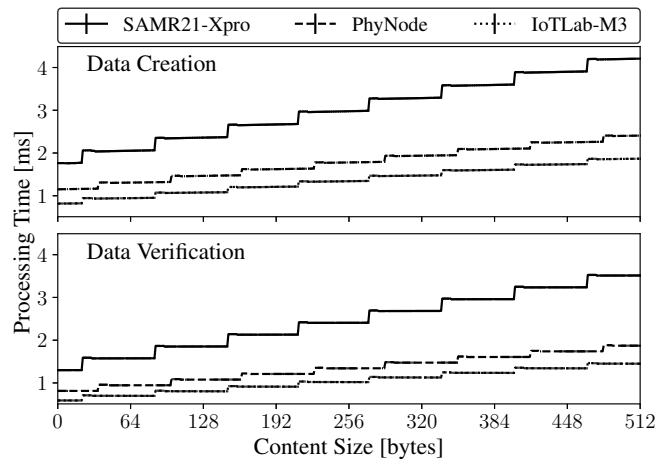}}
    \caption{Computational efforts for signing and verifying data with HMAC(SHA-256) on typical IoT nodes.}
    \label{fig:hmac_gen}
\end{figure}

The advantages of content object security in ICN comes at the price of signing resp. verifying every content chunk that traverses the network. In CCN/NDN, content signatures are usually generated from lightweight crypto hashes. In detail, each content chunk is hashed by SHA-256 followed by a keyed-hash message authentication code (HMAC). This message authentication is provided with our RIOT \cite{bghkl-rosos-18} version of CCN-lite, and we evaluated its performance in benchmarks on common IoT nodes. Figure \ref{fig:hmac_gen} displays the runtime performance as a function of content size for three different IoT boards (running ARM Cortex M0, M3, and M4). Strikingly, the cost of few milliseconds per chunk is fully compliant with networking at the constrained nodes,  which can  send or receive a few packets per second at most. Limitations may derive from energy constraints, though. However, it is safe to conclude that signing and verifying of content is largely compliant to the constrained IoT.

\begin{figure}
  \pgfplotstableread[columns/Name/.style={string type}]{
% SW = Short Weierstrass
% TEE = Twisted Edwads extended
method      SW              TEE
extract     122511580       100265910
sign        122543000       104737590
verify      389525330       307320620
}\dataset
\begin{tikzpicture}
\sisetup{
exponent-product=\cdot
 }
\begin{axis}[
	xbar,
	width=8.5cm,
	height=4.5cm,
	xmin=0,
	xmax=500000000,
	xlabel={CPU time [cycles $\cdot 10^{6}$]},
	xmajorgrids,
	xtick={0, 100000000, 200000000, 300000000, 400000000, 500000000},
	xticklabels={$0$, $100$, $200$, $300$, $400$, $500$},
	scaled x ticks = false,
	yticklabels = {\strut Extract, \strut Sign, \strut Verify},
	ytick=data,
	ymin=-0.5,
	ymax=2.5,
	mark size=4,
	legend entries={Short Weierstrass, Twisted Edwards Extended},
	legend columns=2,
	legend style={draw=none, at={(0.5,1)},xshift=-0.5cm, anchor=south},
	reverse legend,
	nodes near coords={
		\pgfmathfloattofixed{\pgfplotspointmeta} % Convert floating point to fixed point
		\num[
%			scientific-notation = fixed,
%			fixed-exponent = 6,
%			exponent-base = 0,
%			round-mode = places,
%			round-precision = 1
		]{\mult{0.000001}{\pgfmathresult}}},
	every node near coord/.append style={ black }
	]
\addplot[white, draw=black, pattern=dots, area legend]           table[x=SW, y expr=\coordindex] \dataset;
\addplot[white, draw=black, pattern=vertical lines, area legend] table[x=TEE, y expr=\coordindex] \dataset;
,\end{axis}
\end{tikzpicture}
  \caption{Performance of identity-based elliptic curve crypto\-graphy on a PhyNode.}
  \label{fig:iot_vbnn_ibs_performance}
\end{figure}
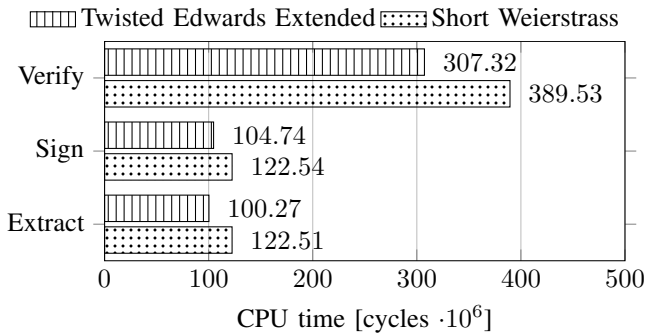

HMAC runs with a pre-established secret, which in an automated environment requires a key management scheme. We devised a key distribution mechanism using identity-based cryptography that operates on elliptic curves. In detail, we implemented the twisted Edwards Curve 25519 in the Relic library on RIOT and compared with an existing short Weierstrass ECC on the Cortex M4 board running at 168 MHz. Figure \ref{fig:iot_vbnn_ibs_performance} shows the runtime results for signature generation and verification of the key establishment that {\em needs to be performed only once}. Clearly, these asymmetric crypto operations are very expensive on our weak microcontroller with  runtimes in the order of minutes. However, they are feasible and enable powerful schemes for autoconfiguration and self-management. Alternative schemes of lower complexity also exist.

\section{Conclusion and Outlook}
\label{sec:c+o}

The industrial IoT connects safety critical environments to the Internet, requiring a high level of reliability and security for data, infrastructure, and end devices.
Multiple stakeholders in this inter-domain communication challenge security, but current protocols in the IoT are weak in meeting these demands.

In this paper, we start from a real-world use  case and derive a security perspective for an information-centric industrial Internet of Things.
We argue three observations.
First, data should be secured intrinsically, with respect to integrity and secrecy so that it can be transparently distributed and stored by any node in the network.
Second, low-end devices as deployed in the IoT should be secured from unsolicited traffic to preserve resources such as battery power and processing.
Third, the delivery infrastructure  requires dedicated protection to increase data availability.
ICN, which abandons the end-to-end paradigm and provides in-network caching, overcomes common attack vectors in the current Internet. 

In future work, real-world deployment and experimentation is needed to evaluate and harden the contributions ICN can make towards a safe and secure industrial Internet of Things.

\balance
\begin{small}
\bibliographystyle{IEEEtran}
\bibliography{main_arxiv.bbl}

% Generated by IEEEtran.bst, version: 1.14 (2015/08/26)
\begin{thebibliography}{10}
\providecommand{\url}[1]{#1}
\csname url@samestyle\endcsname
\providecommand{\newblock}{\relax}
\providecommand{\bibinfo}[2]{#2}
\providecommand{\BIBentrySTDinterwordspacing}{\spaceskip=0pt\relax}
\providecommand{\BIBentryALTinterwordstretchfactor}{4}
\providecommand{\BIBentryALTinterwordspacing}{\spaceskip=\fontdimen2\font plus
\BIBentryALTinterwordstretchfactor\fontdimen3\font minus
  \fontdimen4\font\relax}
\providecommand{\BIBforeignlanguage}[2]{{%
\expandafter\ifx\csname l@#1\endcsname\relax
\typeout{** WARNING: IEEEtran.bst: No hyphenation pattern has been}%
\typeout{** loaded for the language `#1'. Using the pattern for}%
\typeout{** the default language instead.}%
\else
\language=\csname l@#1\endcsname
\fi
#2}}
\providecommand{\BIBdecl}{\relax}
\BIBdecl

\bibitem{adiko-sind-12}
B.~Ahlgren, C.~Dannewitz, C.~Imbrenda, D.~Kutscher, and B.~Ohlman, ``{A Survey
  of Information-Centric Networking},'' \emph{IEEE Communications Magazine},
  vol.~50, no.~7, pp. 26--36, July 2012.

\bibitem{jstp-nnc-09}
V.~Jacobson, D.~K. Smetters, J.~D. Thornton, and M.~F. Plass, ``{Networking
  Named Content},'' in \emph{5th Int. Conf. on emerging Networking Experiments
  and Technologies (ACM CoNEXT'09)}.\hskip 1em plus 0.5em minus 0.4em\relax New
  York, NY, USA: ACM, Dec. 2009, pp. 1--12.

\bibitem{olg-ccnte-10}
S.~Y. Oh, D.~Lau, and M.~Gerla, ``{Content Centric Networking in tactical and
  emergency MANETs},'' in \emph{2010 IFIP Wireless Days}.\hskip 1em plus 0.5em
  minus 0.4em\relax Piscataway, NJ, USA: IEEE, Oct 2010, pp. 1--5.

\bibitem{bmhsw-icnie-14}
\BIBentryALTinterwordspacing
E.~Baccelli, C.~Mehlis, O.~Hahm, T.~C. Schmidt, and M.~W{\"a}hlisch,
  ``{Information Centric Networking in the IoT: Experiments with NDN in the
  Wild},'' in \emph{Proc. of 1st ACM Conf. on Information-Centric Networking
  (ICN-2014)}.\hskip 1em plus 0.5em minus 0.4em\relax New York: ACM, September
  2014, pp. 77--86. [Online]. Available:
  \url{http://dx.doi.org/10.1145/2660129.2660144}
\BIBentrySTDinterwordspacing

\bibitem{sblwy-ndnti-16}
W.~Shang, A.~Bannis, T.~Liang, Z.~Wang, Y.~Yu, A.~Afanasyev, J.~Thompson,
  J.~Burke, B.~Zhang, and L.~Zhang, ``{Named Data Networking of Things (Invited
  Paper)},'' in \emph{Proc. of IEEE International Conf. on Internet-of-Things
  Design and Implementation (IoTDI)}.\hskip 1em plus 0.5em minus 0.4em\relax
  Los Alamitos, CA, USA: IEEE Computer Society, 2016, pp. 117--128.

\bibitem{szsmb-avdir-17}
E.~M. Schooler, D.~Zage, J.~Sedayao, H.~Moustafa, A.~Brown, and M.~Ambrosin,
  ``{An Architectural Vision for a Data-Centric IoT: Rethinking Things, Trust
  and Clouds},'' in \emph{IEEE 37th Intern. Conference on Distributed Computing
  Systems (ICDCS)}.\hskip 1em plus 0.5em minus 0.4em\relax Piscataway, NJ, USA:
  IEEE, June 2017, pp. 1717--1728.

\bibitem{pf-britu-15}
G.~C. Polyzos and N.~Fotiou, ``{Building a reliable Internet of Things using
  Information-Centric Networking},'' \emph{Journal of Reliable Intelligent
  Environments}, vol.~1, no.~1, pp. 47--58, 2015.

\bibitem{abcmr-inmcd-16}
M.~Amadeo, O.~Briante, C.~Campolo, A.~Molinaro, and G.~Ruggeri,
  ``{Information-centric networking for M2M communications: Design and
  deployment},'' \emph{Computer Communications}, vol. 89--90, pp. 105 -- 116,
  2016.

\bibitem{mwt-tucin-16}
B.~Mathieu, C.~Westphal, and P.~Truong, ``Towards the usage of ccn for iot
  networks,'' in \emph{Internet of Things (IoT) in 5G Mobile
  Technologies}.\hskip 1em plus 0.5em minus 0.4em\relax Cham, Switzerland:
  Springer, 2016, pp. 3--24.

\bibitem{bhgws-rotoi-13}
E.~Baccelli, O.~Hahm, M.~G{\"u}nes, M.~W{\"a}hlisch, and T.~C. Schmidt, ``{RIOT
  OS: Towards an OS for the Internet of Things},'' in \emph{Proc. of the 32nd
  IEEE INFOCOM. Poster}.\hskip 1em plus 0.5em minus 0.4em\relax Piscataway, NJ,
  USA: IEEE Press, 2013, pp. 79--80.

\bibitem{dgv-clfos-04}
A.~Dunkels, B.~Gr{\"o}nvall, and T.~Voigt, ``{Contiki - A Lightweight and
  Flexible Operating System for Tiny Networked Sensors.}'' in \emph{{Proc. of
  IEEE Local Computer Networks (LCN)}}.\hskip 1em plus 0.5em minus 0.4em\relax
  IEEE Computer Society, 2004, pp. 455--462.

\bibitem{saz-dinps-16}
W.~Shang, A.~Afanasyev, and L.~Zhang, ``{The Design and Implementation of the
  NDN Protocol Stack for RIOT-OS},'' in \emph{Proc. of IEEE GLOBECOM
  2016}.\hskip 1em plus 0.5em minus 0.4em\relax Washington, DC, USA: IEEE,
  2016, pp. 1--6.

\bibitem{bjkrb-ssnuc-19}
H.~Bornholdt, D.~Jost, P.~Kisters, et al., ``{SANE: Smart Networks for Urban Citizen   Participation},'' in \emph{2019 26th International Conference on
  Telecommunications (ICT) (ICT 2019)}.\hskip 1em plus 0.5em minus 0.4em\relax
  Piscataway, NJ, USA: IEEE Press, April 2019.

\bibitem{gksw-hrrpi-18}
\BIBentryALTinterwordspacing
C.~G{\"u}ndogan, P.~Kietzmann, T.~C. Schmidt, and M.~W{\"a}hlisch, ``{HoPP:
  Robust and Resilient Publish-Subscribe for an Information-Centric Internet of
  Things},'' in \emph{Proc. of the 43rd IEEE Conference on Local Computer
  Networks (LCN)}.\hskip 1em plus 0.5em minus 0.4em\relax Piscataway, NJ, USA:
  IEEE Press, Oct. 2018, pp. 331 -- 334. [Online]. Available:
  \url{http://doi.org/10.1109/LCN.2018.8638030}
\BIBentrySTDinterwordspacing

\bibitem{mqtt311}
\BIBentryALTinterwordspacing
A.~Banks and R.~G. (Eds.), ``{MQTT Version 3.1.1},'' OASIS, OASIS Standard,
  October 2014. [Online]. Available:
  \url{http://docs.oasis-open.org/mqtt/mqtt/v3.1.1/os/mqtt-v3.1.1-os.html}
\BIBentrySTDinterwordspacing

\bibitem{RFC-7252}
Z.~Shelby, K.~Hartke, and C.~Bormann, ``{The Constrained Application Protocol
  (CoAP)},'' IETF, RFC 7252, June 2014.

\bibitem{wsv-bipmc-12}
\BIBentryALTinterwordspacing
M.~W{\"a}hlisch, T.~C. Schmidt, and M.~Vahlenkamp, ``{Bulk of Interest:
  Performance Measurement of Content-Centric Routing},'' in \emph{Proc. of ACM
  SIGCOMM, Poster Session}.\hskip 1em plus 0.5em minus 0.4em\relax New York:
  ACM, August 2012, pp. 99--100. [Online]. Available:
  \url{http://conferences.sigcomm.org/sigcomm/2012/paper/sigcomm/p99.pdf}
\BIBentrySTDinterwordspacing

\bibitem{gtuz-ddndn-13}
P.~Gasti, G.~Tsudik, E.~Uzun, and L.~Zhang, ``{DoS and DDoS in Named Data
  Networking},'' in \emph{Proc. of ICCCN}.\hskip 1em plus 0.5em minus
  0.4em\relax {IEEE}, 2013, pp. 1--7.

\bibitem{wsv-bdpts-13}
\BIBentryALTinterwordspacing
M.~W{\"a}hlisch, T.~C. Schmidt, and M.~Vahlenkamp, ``{Backscatter from the Data
  Plane -- Threats to Stability and Security in Information-Centric Network
  Infrastructure},'' \emph{Computer Networks}, vol.~57, no.~16, pp. 3192--3206,
  Nov. 2013. [Online]. Available:
  \url{http://dx.doi.org/10.1016/j.comnet.2013.07.009}
\BIBentrySTDinterwordspacing

\bibitem{sws-rcani-15}
S.~Al-Sheikh, M.~W{\"a}hlisch, and T.~C. Schmidt, ``{Revisiting Countermeasures
  Against NDN Interest Flooding},'' in \emph{2nd ACM Conference on
  Information-Centric Networking, Poster Session}, ser. ICN 2015.\hskip 1em
  plus 0.5em minus 0.4em\relax New York: ACM, Oct. 2015, pp. 195--196.

\bibitem{gklp-ncmcm-18}
\BIBentryALTinterwordspacing
C.~G{\"u}ndogan, P.~Kietzmann, M.~Lenders, H.~Petersen, T.~C. Schmidt, and
  M.~W{\"a}hlisch, ``{NDN, CoAP, and MQTT: A Comparative Measurement Study in
  the IoT},'' in \emph{Proc. of 5th ACM Conference on Information-Centric
  Networking (ICN)}.\hskip 1em plus 0.5em minus 0.4em\relax New York, NY, USA:
  ACM, September 2018. [Online]. Available:
  \url{https://conferences.sigcomm.org/acm-icn/2018/proceedings/icn18-final46.pdf}
\BIBentrySTDinterwordspacing

\bibitem{mqttsn12}
\BIBentryALTinterwordspacing
A.~Stanford-Clark and H.~L. Truong, ``{MQTT For Sensor Networks (MQTT-SN)
  Version 1.2},'' IBM, Protocol Specification, November 2013. [Online].
  Available:
  \url{http://mqtt.org/new/wp-content/uploads/2009/06/MQTT-SN\_spec\_v1.2.pdf}
\BIBentrySTDinterwordspacing

\bibitem{bghkl-rosos-18}
\BIBentryALTinterwordspacing
E.~Baccelli, C.~G{\"u}ndogan, O.~Hahm, P.~Kietzmann, M.~Lenders, H.~Petersen,
  K.~Schleiser, T.~C. Schmidt, and M.~W{\"a}hlisch, ``{RIOT: an Open Source
  Operating System for Low-end Embedded Devices in the IoT},'' \emph{IEEE
  Internet of Things Journal}, vol.~5, no.~6, pp. 4428--4440, December 2018.
  [Online]. Available: \url{http://dx.doi.org/10.1109/JIOT.2018.2815038}
\BIBentrySTDinterwordspacing

\end{thebibliography}
\end{small}

\end{document}